\documentclass[aps,prl,10pt,twocolumn,superscriptaddress,showpacs,preprintnumbers,nofootinbib]{revtex4-1}
\usepackage{titlesec}
\usepackage{hyperref}
\usepackage{graphicx}
\usepackage{amsfonts,amsmath,amssymb,bm,bbm}

\hypersetup{
    pdfnewwindow=true,      
    colorlinks=true,       
    linkcolor=blue,          
    citecolor=blue,        
    filecolor=blue,      
    urlcolor=blue        
}

\titleformat{\section}[runin]
  {\normalfont\itshape}{\thesection}{1em}{}[.---]
\titlespacing*{\section}{0.5cm}{2em}{1em}
\titleformat{\subsection}[runin]
  {\normalfont\itshape}{\thesubsection}{1em}{}[---]
\titlespacing*{\subsection}{1cm}{2em}{1em}

\begin{document}


\title{Constraints on neutrino speed, weak equivalence principle violation, Lorentz invariance violation, and dual lensing from the first high-energy astrophysical neutrino source TXS 0506+056}

\author{Ranjan Laha}
\affiliation{PRISMA Cluster of Excellence and
             Mainz Institute for Theoretical Physics,
             Johannes Gutenberg-Universit\"{a}t Mainz, 55099 Mainz, Germany} 
\affiliation{Theoretical Physics Department, CERN, 1211 Geneva, Switzerland \\
{\tt  \href{mailto:ranjan.laha@cern.ch}{ranjan.laha@cern.ch}}
{\tt \footnotesize \href{http://orcid.org/0000-0001-7104-5730}{0000-0001-7104-5730} \smallskip}}

\date{\today}

\begin{abstract}
We derive stringent constraints on neutrino speed, weak equivalence principle violation, Lorentz invariance violation, and dual lensing from the first high-energy astrophysical neutrino source: TXS 0506+056.  Observation of neutrino (IceCube-170922A) and photons in a similar time frame and from the same direction is used to derive these limits.  We describe ways in which these constraints can be further improved by orders of magnitude.
\end{abstract}

\maketitle

\section{Introduction}
\label{sec:introduction}

IceCube, along with other electromagnetic telescopes, have discovered the first high-energy astrophysical neutrino source\,\cite{IceCube:2018dnn, IceCube:2018cha}.  Predictably, a lot of excitement has been generated by this momentous discovery\,\cite{Gao:2018mnu, Righi:2018hhu, Ahnen:2018mvi, Albert:2018kjg, Cerruti:2018tmc, Padovani:2018acg, Keivani:2018rnh, Abeysekara:2018oub, Murase:2018iyl}.  Together with the prior discovery of a diffuse flux of high-energy astrophysical neutrinos\,\cite{Aartsen:2013bka, Aartsen:2016xlq}, these observations provide us a new way to probe beyond the Standard model physics\,\cite{Feldstein:2013kka, Murase:2015gea,  Bhattacharya:2014yha, Kopp:2015bfa, Dey:2015eaa, Bhattacharya:2016tma,  Dasgupta:2012bd, Vogel:2017fmc, Bustamante:2017xuy, Dhuria:2017ihq}, new IceCube signal morphologies\,\cite{Kistler:2016ask, Li:2016kra}, and atmospheric prompt neutrinos\,\cite{Aartsen:2015nss, Bhattacharya:2016jce, Gauld:2015kvh, Garzelli:2016xmx, Halzen:2016thi, Laha:2016dri, Goncalves:2017lvq, Laha:2013eev}.    

IceCube detected a through-going muon with a deposited energy of 23.7 $\pm$ 2.8 TeV on September 22$^{\rm nd}$ 2017.  Depending on the underlying astrophysical neutrino spectrum, the most probable energy of the parent neutrino (IceCube-170922A) is 290 TeV or 311 TeV, and the 90\% C.L. energy interval is [183 TeV, 4.3 PeV] or [200 TeV, 7.5 PeV].  The best fit direction of this event is R.A. = 77.43 $^{+0.95}_{-0.65}$ degrees and Dec = 5.72 $^{+0.50}_{-0.30}$ degrees, and it is $\sim$ 0.1$^{\rm o}$ from the blazar TXS 0506+056 situated at $z$ = 0.3365\,\cite{IceCube:2018dnn, IceCube:2018cha, Paiano:2018qeq}.  The significance of this single event is at the level of 3$\sigma$.  The case for TXS 0506+056 being the first high-energy astrophysical neutrino source is further bolstered by the fact that a $\sim$ 3.5$\sigma$ evidence of neutrino emission from the same direction is observed in prior data\,\cite{IceCube:2018cha}.  The IceCube collaboration did not present a combined significance of these two observations.  A naive combination of them yields $\sim$ 4.6$\sigma$ which underscores the importance of this detection.

The detection of neutrinos and photons is utilized in this work to derive new constraints on neutrino speed, weak equivalence principle violation, Lorentz invariance violation, and dual lensing phenomenon.  Numerous studies have been undertaken in the first 3 topics, using SN 1987A and other multi-messenger observations\,\cite{Longo:1987ub, Longo:1987gc, Krauss:1987me, Pakvasa:1988gd, LoSecco:1988jg, Stecker:2014oxa, Wei:2015hwd, Wei:2016mzj, Wu:2016igi, Wang:2016lne, Wei:2016ygk, Nusser:2016wzr, Desai:2016nqu,  Wei:2017nyl, Bertolami:2017opd, Wei:2019vzw}.  The last topic has been studied recently by physicists interested in quantum gravity\,\cite{Freidel:2011mt, AmelinoCamelia:2011gy, Loret:2012zz, AmelinoCamelia:2012rz, Amelino-Camelia:2017pne}, but application to astroparticle phenomenology is few\,\cite{Amelino-Camelia:2016wpo}.

\begin{table}[b]

\caption{Summary of results derived in this work.  Here WEP stands for weak equivalence principle, LI denotes Lorentz invariance, and $E_{\rm Pl}$ is the Planck energy.  We display the explicit dependence of the constraints on the time difference between neutrino and photon detection, $\Delta t$.  Due to the continuous photon emission, we are forced to assume a value of $\Delta t$ to derive our numerical for the first four cases displayed below.  In all the relevant cases, we show the explicit scaling of our results with $\left(\Delta t/{\rm 7 \, days}\,\right)$.} 

\begin{ruledtabular}

\begin{tabular}{lcc}
Scenario & Relevant parameter & Limit \\ 
\hline
Neutrino speed & $\dfrac{|v - c|}{c}$ & $\lesssim 4.2 \times 10^{-12} \, \left(\Delta t/{\rm 7 \, days}\,\right)$  \\
WEP violation & $\gamma_\nu - \gamma_{\rm ph}$ & $\lesssim 3.5 \times 10^{-7}  \, \left(\Delta t/ {\rm 7 \, days}\,\right)$ \\
LI violation & $E_{\rm QG, 1}$ & $\gtrsim 6 \times 10^{-3}$ $E_{\rm Pl} \, \left(\Delta t/{\rm 7 \, days}\,\right)^{-1}$ \\
LI violation & $E_{\rm QG, 1}$ & $\gtrsim 1.6 \times 10^{-8}$ $E_{\rm Pl} \, \left(\Delta t/{\rm 7 \, days}\,\right)^{-1/2}$ \\
Dual lensing & $k_{\rm d.l.}$ & $\lesssim 7.3 \times 10^{10}$ \\

\end{tabular}

\end{ruledtabular}
\label{tab: results}
\end{table}

Time delay between neutrinos and photons from distant astrophysical sources provides one of the most stringent bounds on neutrino velocity, and can even constrain neutrino mass in the near future\,\cite{Kyutoku:2017wnb}.  One of the main hypotheses of general relativity states that the motion of various neutral massless particles depends on the intervening gravitational potential\,\cite{Will:2005va, Will:2014kxa}.  The search for the violation of Einstein's weak equivalence principle (mentioned above) is one of the main ways to search for avenues beyond general relativity.  Some quantum gravity models predict Lorentz invariance violation at some energy scale $E_{\rm QG}$\,\cite{AmelinoCamelia:1997gz}.  Various tests of Lorentz invariance violation have been proposed and executed\,\cite{Ackermann:2009aa, Vasileiou:2013vra, Ellis:2008fc, Jacob:2006gn}.  The phenomenon of dual lensing implies that neutral messengers will diverge from their original paths and will appear as arriving from different directions to a distant observer.  We point out that a close match in the direction of the neutrino IceCube-170922A with TXS 0506+056 implies a severe constraint on the coefficient of proportionality of this phenomenon.  We use the cosmological parameters ($\Omega_m$, $\Omega_\Lambda$, and $H_0$) measured by the Planck collaboration\,\cite{Ade:2015xua}.  Our results are summarized in Table\,\ref{tab: results}.

Our work is inspired by various earlier studies which have been performed using the multi-messenger observations of SN 1987A\,\cite{Longo:1987ub, Longo:1987gc, Krauss:1987me, Pakvasa:1988gd, LoSecco:1988jg}.  However, there are important differences between the observation of SN 1987A and multi-messenger observations of TXS 0506+056.  In the case of SN 1987A, no gamma-ray emission was observed before the neutrino emission.  In other words, the neutrino observation was the ``start time trigger" for multi-messenger observations.  The case with multi-messenger observations of TXS 0506+056 is very different: this blazar was a well-known gamma-ray source even before the neutrino observation.  Although, the blazar is a continuous gamma-ray emitter, Fermi-LAT had detected an enhanced emission at GeV energies from April 2017.  The neutrino (IceCube-170922A) was detected during the period of enhanced gamma-ray activity\,\cite{IceCube:2018dnn, IceCube:2018cha}.  Enhanced emission was also observed at lower photon energies before and during the neutrino detection period.  Theoretical modeling has indicated that the higher energy gamma-rays detected from TXS 0506+056 is most probably leptonic in origin, and that the pionic gamma-rays are absorbed and emitted at lower energies\,\cite{Gao:2018mnu, Keivani:2018rnh, Liu:2018utd, Reimer:2018vvw, Sahakyan:2018voh, He:2018snd, Cerruti:2018tmc, Xue:2019txw, Rodrigues:2018tku, Halzen:2018iak, Banik:2019jlm, Banik:2019twt}.  As such, it becomes model dependent to find the time difference between the neutrino and the photon emission.  This choice of a time difference between the photon and the neutrino emission is the largest uncertainty in our deduced limits.  The choice of the time difference was fairly model-independent for SN 1987A, whereas we are forced to choose a time difference in this scenario.  We display our results in a way which transparently displays our choice of the time difference.  In order to obtain numerical values of the relevant constraints, we assume $\Delta t$ = 7 days.  We remind the reader that this choice of 7 days is an assumption and does not have a physical significance.  In order to make our results easily applicable and comparable to specific models in which one can deduce the time difference, we quote all our relevant results such that it prominently shows the dependence on the fraction ($\Delta t$/7 days).    
 
\section{Constraint on neutrino speed}
\label{sec: neutrino speed}

Targeted high-energy gamma-ray observations of TXS 0506+056 in the days following IceCube-170922A indicated that the blazar was already undergoing a flare\,\cite{IceCube:2018dnn}.  Fermi-LAT and AGILE detected gamma-ray emission when the observations were binned into 7 days and 13 days respectively.  A smaller value of binning (even sub second) is expected during the observation of gamma-ray bursts in photons and neutrinos, and this can improve the following constraints by orders of magnitude.  It is worthwhile to mention that this source was already detected via gamma-rays in the time prior to the neutrino detection\,\cite{IceCube:2018dnn}.  Theoretical calculations have shown that the detected neutrinos and high-energy gamma-rays do not originate from the same interaction\,\cite{Gao:2018mnu, Keivani:2018rnh, Liu:2018utd, Reimer:2018vvw, Sahakyan:2018voh, He:2018snd, Cerruti:2018tmc, Xue:2019txw, Rodrigues:2018tku, Halzen:2018iak, Banik:2019jlm, Banik:2019twt}.  The time delay, $\Delta t$, which is the difference between the time travel of neutrinos and photons from the source to the detector is obviously more precisely determined if one detects neutrino and the photons from the same interaction.  The current experimental data do not give us an opportunity to apply such a  tight constraint.  In the absence of near co-incident data, we are forced to assume a value of $\Delta t$.  Since we scale our limits with the factor ($\Delta t/7{\rm \,days}$), one can easily estimate how stringent the limit will be for the case of a near coincident detection.  

We also note the coincidence of the neutrino event IceCube-160731 with the astrophysical source AGL J1418+0008\,\cite{Lucarelli:2017hhh}.  In this case, the significance is higher and the gamma-rays were detected within one or two days before the neutrino detection\footnote{We thank the referee for pointing out this observation}.  The techniques presented in this work can also be applied to this detection and we encourage future work in this direction.      

The constraint on relative velocity difference between neutrinos and photons is\,\cite{Schaefer:1998zg}
\begin{eqnarray}
\dfrac{|v - c|}{c} \leq \dfrac{c \, \Delta t}{\dfrac{c}{H_0} \int_0 ^z \dfrac{dz'}{\sqrt{\Omega_\Lambda + \Omega_m (1+z')^3}}} \, ,
\label{eq: neutrino speed equation}
\end{eqnarray}  
where the denominator is the co-moving distance to the source.  Using the values mentioned above, we find
\begin{eqnarray}
\dfrac{|v - c|}{c} \lesssim 4.2 \times 10^{-12} \, \left(\dfrac{\Delta t}{\rm 7 \, days}\,\right) ,
\label{eq: neutrino speed value}
\end{eqnarray}  
which is 3 orders of magnitude and 6 orders of magnitude more stringent compared to that obtained from SN 1987A observations and laboratory measurements, respectively\,\cite{Longo:1987ub, Adamson:2015ayc, AlvarezSanchez:2012wg, Antonello:2012hg, Adam:2011faa}, assuming $\Delta t$ = 7 days.  We also show the scaling of our result with $\Delta t$ so that the reader can quickly estimate how much these limits can improve if one observes a neutrino and a photon with a smaller time delay. 
  
If Lorentz violation is implemented by an effective operator, then one can translate this limit on the neutrino speed to the corresponding charged lepton speed.  There has been various estimates on the charged lepton speed using different observations (see Ref.\,\cite{Giudice:2011mm} for a compilation).  Given the energy of the neutrino, this limit can also be translated into a limit on the speed of charged leptons at few hundreds of TeV, where no current limits exist\,\cite{Giudice:2011mm}.  Given that the energy of this neutrino is much larger than the W-boson mass, the corresponding limit on the charged lepton speed depends on the energy dependence of the $\delta c \equiv |v - c|$.  However, it is not guaranteed that Lorentz violation can always be described by an effective operator.  This situation is realized in space-time foam models where particle interactions are not described by an effective field theory due to the space-time uncertainty effects\,\cite{Ellis:2008gg, Li:2009tt}.  Given these model uncertainties, we prefer to not study the limits on the charged lepton speed in more detail.

\section{Constraint on weak equivalence principle}
\label{sec: weak equivalence principle}

As first derived by Shapiro, a neutral particle experiences a time delay while traveling through a gravitational potential $U(r)$\,\cite{Shapiro:1964uw}
\begin{eqnarray}
\delta t = - \dfrac{1 + \gamma}{c^3} \int U(r) \, dr \,,
\label{eq: Shapiro time delay}
\end{eqnarray}
where $\gamma$ = 1 in general relativity.  Weak equivalence principle violation implies that the value of $\gamma$ for neutrino is not equal to that of photon, i.e., $\gamma_\nu \neq \gamma_{\rm ph}$.  This inequality results in an arrival time delay between neutrinos and photons which are emitted simultaneously:
\begin{eqnarray}
\Delta t =  \dfrac{\gamma_\nu - \gamma_{\rm ph}}{c^3} \int U(r) \, dr \,.
\label{eq: Shapiro time delay neutrinos and photon}
\end{eqnarray}
For extragalactic sources, it has been shown that the gravitational potential is dominated by the Laniakea supercluster\,\cite{Wei:2016ygk, Tully:2014gfa}.  Assuming a Keplerian potential, we obtain\,\cite{Wei:2016ygk}
\begin{eqnarray}
\Delta t &=&  \dfrac{\gamma_\nu - \gamma_{\rm ph}}{c^3} \, G_N M_L \nonumber\\
&\times& {\rm ln} \, \dfrac{(d + \sqrt{d^2 - b^2}) \, (r_L + s_n \sqrt{r_L^2 - b^2})}{b^2} \, ,
\label{eq: Shapiro time delay Laniakea}
\end{eqnarray}
where the mass of Laniakea supercluster is $M_L \approx 10^{17}$ M$_\odot$, and $d$ denotes the distance to the cosmic source from the Laniakea supercluster.  We can well approximate this distance to be that between the Laniakea supercluster and the Earth.  The impact parameter ($b$) depends on the distance from the Earth to the Laniakea center ($r_L = 79$ Mpc), the coordinates of the source in R.A. and Dec., ($\beta_s$, $\delta_s$), and the coordinate of the Laniakea center, $\beta_L = 158^{\rm o}$ and $\delta_L = - 46^{\rm o}$\,\cite{Wei:2016ygk}
\begin{eqnarray}
b = r_L \, \sqrt{1-({\rm cos}\, \delta_s \, {\rm cos}\, \delta_L \, {\rm cos}\, (\beta_s - \beta_L) + {\rm sin}\, \delta_s \, {\rm sin}\, \delta_L )^2} \,. \phantom{111}
\label{eq: impact parameter}
\end{eqnarray}
The value of $s_n = \pm$1 depends on whether the source is located in the same direction (+ve sign) or opposite (-ve sign) w.r.t. to the Lanikea center.

Using the data obtained from the neutrino emission of TXS 0506+056, we obtain
\begin{eqnarray}
\gamma_\nu - \gamma_{\rm ph} \lesssim 3.5 \times 10^{-7} \, \left(\dfrac{\Delta t}{\rm 7 \, days}\,\right).
\label{eq: Einstein WEP value}
\end{eqnarray}  
Assuming such a time delay, this constraint is orders of magnitude stronger than that derived using the observation of SN 1987A.  Besides using a smaller time interval measurement, the other ways to improve this constraint is to identify the supercluster in which the source is located and to estimate the contribution due to the large scale structure.

\section{Constraint on Lorentz invariance violation}
\label{sec: Lorentz invariance violation}

Quantum gravity models which postulate Lorentz invariance violation imply that there is a modification in the energy ($E$) - momentum ($p$) dispersion relationship for a particle of mass $m$:
\begin{eqnarray}
E^2 = p^2 + m^2 \pm E^2 \Bigg( \dfrac{E}{E_{\rm QG}} \Bigg)^n \, ,
\label{eq: energy momentum dispersion relation}
\end{eqnarray}
where the $\pm$ sign correspond to subluminal or superluminal propagation.  Typically only the index $n = 1, 2$ are considered and we follow the same prescription for the constraints presented here.  For cosmological propagation, this implies that the time delay due to Lorentz invariance violation when the energy of one particle is much larger than the other particle is\,\cite{Wang:2016lne}
\begin{eqnarray}
\Delta t = \dfrac{1 + n}{2 H_0} \, \Bigg(\dfrac{E}{E_{\rm QG}} \Bigg)^n \, \int _0 ^z \dfrac{(1+z')^n}{\sqrt{\Omega_\Lambda + \Omega_M (1+z')^3}} \, dz' \, . \phantom{11}
\label{eq: time delay due to Lorentz invariance violation}
\end{eqnarray}
We obtain for linear $(n = 1)$ Lorentz invariance violation 
\begin{eqnarray}
E_{\rm QG, 1} \gtrsim 6 \times 10^{-3} \, E_{\rm Pl} \,  \left(\dfrac{\Delta t}{\rm 7 \, days}\,\right)^{-1},
\label{eq: linear Lorentz invariance violation limit}
\end{eqnarray}
and for quadratic $(n = 2)$ Lorentz invariance violation
\begin{eqnarray}
E_{\rm QG, 2} \gtrsim 1.6 \times 10^{-8} \, E_{\rm Pl} \,\left(\dfrac{\Delta t}{\rm 7 \, days}\,\right)^{-1/2},
\label{eq: quadratic Lorentz invariance violation limit}
\end{eqnarray}
where the Planck energy scale is $E_{\rm Pl}$ = 1.22 $\times$ 10$^{19}$ GeV.  For our chosen time delay, these limits are orders of magnitude more stringent than those set by the observations of SN 1987A\,\cite{ Ellis:2008fc}.  The constraint on $E_{\rm QG, 1}$ set via observation of gamma-ray burst is stronger than the limits presented here by about 2 -- 3 orders of magnitude, whereas the constraint derived here on $E_{\rm QG, 2}$ is stronger by a factor of 1.6\,\cite{Ackermann:2009aa, Vasileiou:2013vra, HESS:2011aa}, however, this conclusion assumes that Lorentz invariance violation is present in the photon sector.  Assuming that such a Lorentz invariance violation is only present in the neutrino sector, we have derived the most stringent limits.

\section{Constraint on dual lensing}
\label{sec: Dual lensing}

If the phenomenon of dual lensing is present in Nature, then a distant observer will detect two messengers from different parts of the sky even if they originated from the same astrophysical source.  It is similar to gravitational lensing, although the origins of this effect is beyond beyond the realms of general theory of relativity.  Momentum-space curvature which is hypothesized in certain ways to quantize gravity predicts the phenomena of dual lensing.  A phenomenological approach to dual lensing was undertaken in Ref.\,\cite{Amelino-Camelia:2016wpo} where the authors assumed an energy-independent dual lensing angle for the exploratory study.  However, theoretical studies about this phenomena predict an energy-dependent dual lensing angle\,\cite{Freidel:2011mt, AmelinoCamelia:2011gy, Loret:2012zz, AmelinoCamelia:2012rz, Amelino-Camelia:2017pne}.   

Various theoretical approaches to quantum gravity predict that the dual lensing angle, $\theta_{\rm d.l.}$, between two astrophysical messengers of energies $E_1$ and $E_2$ is proportional to either the sum of difference between the energies, i.e., $\theta_{\rm d.l.} \propto (E_1 \pm E_2)/ E_{\rm Pl}$\,\cite{Freidel:2011mt, AmelinoCamelia:2011gy, Loret:2012zz, AmelinoCamelia:2012rz, Amelino-Camelia:2017pne}.  If this phenomenon is observable only at Planck energy scales, then the constant of proportionality is of order 1.  We define the constant of proportionality, $k_{\rm d.l.}$, and try to constrain it via the neutrino and photon observation of TXS 0506+056:
\begin{eqnarray}
\theta_{\rm d.l.} = k_{\rm d.l.} (E_1 \pm E_2)/ E_{\rm Pl} \, .
\label{eq: dual lensing equation}
\end{eqnarray}

The limit on $k_{\rm d.l.}$ depends on the magnitude and the relative sign between $E_1$ and $E_2$.  We first present the limit on $k_{d.l.}$ assuming the observed neutrino energy and the photon energy.  We will then comment on how the limit can change if one observes a neutrino and a photon with similar energy in the near future.

Given that the observed neutrino energy is much larger than the energy of the observed photons, and that the best-fit neutrino direction is $\sim$ 0.1$^{\rm o}$ from the blazar direction, we find that
\begin{eqnarray}
k_{\rm d.l.} \lesssim 7.3 \times 10^{10} \, .
\label{eq: dual lensing equation}
\end{eqnarray}
In this situation, the limit on $k_{\rm d.l.}$ is almost independent of the sign between $E_1$ and $E_2$ due to the large energy difference between them.    We can define an effective energy scale of dual lensing $E_{\rm eff} \equiv E_{\rm Pl}/ k_{\rm d.l.}$ and the future aim will be to find observables such that $k_{\rm d.l.} \rightarrow$ 1.  If one detects a neutrino and a photon with a similar energy, then the limit on $k_{\rm d.l.}$ will strongly depend on the relative sign between $E_1$ and $E_2$.  Assuming a positive sign between the two energies, it is possible that a higher energy detection will decrease the upper limit on $k_{\rm d.l.}$.  However, a negative sign between $E_1$ and $E_2$ will dramatically weaken the limit.
  
A near future detection of Galactic supernova $\nu_e$ in large water Cherenkov detectors\,\cite{Laha:2013hva, Abe:2016waf, Nikrant:2017nya} or via triangulation\,\cite{Brdar:2018zds} or via detection in large liquid scintillator detectors\,\cite{Laha:2014yua, Fischer:2015oma} can locate the supernova to a few degrees, however, that cannot improve this bound because of the lower energy of supernova neutrinos.  Although electromagnetic observatories can point to a source with sub-degree precision, yet this cannot obtain the best limit due to the much lower energy of photons involved.  A detection of neutrinos produced in the GZK process and the corresponding source or lower energy photon signature can improve this limit by $\sim$ 4 orders of magnitude.     

\section{Conclusion}
\label{sec: Conclusion}
The discovery of the first high-energy astrophysical neutrino source provides a new opportunity to understand various fundamental physical phenomena of the Universe.  The neutrino IceCube-170922A was detected when TXS 0506+056 was in a flaring state and thus the blazar has a high probability to be the birthplace of the neutrino.  An excess of events was also detected from the same direction in prior data.  Both of these observations are at the level of $\gtrsim$ 3$\sigma$ individually and thus TXS 0506+056 is probably the first high-energy astrophysical neutrino source.  Besides a deeper understanding of astrophysics, this discovery will also constrain, or possibly discover, new physics.  We use the detection of neutrinos and photons from TXS 0506+056 to constrain truly exotic physics.  Some of the laws and principle we test in this work are fundamental and thus it is important to experimentally test them in as many different ways as possible.

We summarize our results in Table\,\ref{tab: results}.  The detection of neutrinos and gamma-rays puts a strong constraint on the neutrino velocity.  The same observation also gives a strong constraint on the violation of the weak equivalence principle.  In this case, the gravitational potential of the local supercluster, Laniakea, dominates the measurement.  The identification of the supercluster hosting the astrophysical source and a precise understanding of the contribution of the extragalactic large scale structure can improve this bound.  We use the same observable to constrain the energy scale at which Lorentz violating effects become important.  If Lorentz invariance violation is also present in the photon sector, then the photon observations probe a higher energy scale for the linear form of the violation.  All of these bounds can be improved for a burst-like source where the neutrino and the photon can be detected near co-incidentally.  For a gamma-ray burst, the time delay measurement can be $\lesssim$ 1 second and thus these limits will improve by as much as $\sim$ 6 orders of magnitude.  We also constrain the coefficient of the dual lensing phenomenon via the sub-degree angular distance between IceCube-170922A and TXS 0506+056.  We point out that a near future detection of neutrinos from the GZK process and the subsequent photon observation or source identification can improve this limit by $\sim$ 4 orders of magnitude.  We emphasize that the observed neutrinos and photons from TXS 0506+056 do not arise from the same interaction.  Given this physical insight along with the fact that the blazar was already in a flaring state when the neutrino was detected, we are forced to assume a time difference in order to find our constraints.  We display all our results such that the reader is aware of this assumption of the time difference.  Due to this assumed time difference, it is difficult to compare our limits with those derived from SN 1987A and we do not undertake this comparison.  One can undertake such a comparison if one assumes a model which gives the intrinsic time delay between the neutrino and photon.  It is also to be noted that the statistical significance of this observation is not as high as that of SN 1987A.  Near future discovery of a high-energy astrophysical neutrino sources with a much shorter time difference between neutrino and photon can help us improve the SN 1987A limits by order of magnitude.

We hope to have convinced the reader that the first observation of high-energy neutrinos from an astrophysical source have provided some of the strongest constraints on various exotic physics extensions.  We also point out that there is scope for improvement by orders of magnitude in these type of tests, and one should try to find newer ways to constrain these and other relevant parameters.

{\it Note added:}  After the submission of this paper to arXiv, two papers on related topics appeared online: Refs.\,\cite{Ellis:2018ogq, Boran:2018ypz}.  After this paper appeared on arXiv, we noticed Ref.\,\cite{Wei:2018ajw} on the preprint server.

\section*{Acknowledgments} 
R.L. was supported in Mainz by German Research  Foundation  (DFG)  under  Grant  Nos.  EXC-1098, KO 4820/1-1, FOR 2239, and from the European Research Council (ERC) under the European Union's Horizon 2020 research and innovation programme  (grant  agreement  No.  637506,  ``$\nu$Directions") awarded to Joachim Kopp.  We thank the referee for comments.

\bibliographystyle{kp}
\bibliography{Bibliography/references}	

\end{document}